\documentclass[prl,twocolumn,showpacs,superscriptaddress]{revtex4}
\usepackage{amsmath,amssymb,graphicx,epstopdf,color,bm,soul,upgreek, mathtools}

\begin{document}

\title{Evolution of Temporal Coherence in Confined Exciton-Polariton Condensates}
\author{M. Klaas}
\affiliation{Technische Physik, Wilhelm-Conrad-R\"ontgen-Research Center for Complex
Material Systems, Universit\"at W\"urzburg, Am Hubland, D-97074 W\"urzburg,
Germany}
\author{H. Flayac}
\affiliation{Institute of Theoretical Physics, Ecole Polytechnique Federale de Lausanne (EPFL), CH-1015 Lausanne, Switzerland}
\author{M. Amthor}
\affiliation{Technische Physik, Wilhelm-Conrad-R\"ontgen-Research Center for Complex Material Systems, Universit\"at W\"urzburg, Am Hubland, D-97074 W\"urzburg, Germany}
\author{I.~G.~Savenko}
\affiliation{Center for Theoretical Physics of Complex Systems, Institute for Basic Science (IBS), Daejeon 34051, Republic of Korea.}
\affiliation{Institute of Semiconductor Physics, Siberian Branch of Russian Academy of Sciences, Novosibirsk, 630090, Russia}
\author{S. Brodbeck}
\affiliation{Technische Physik, Wilhelm-Conrad-R\"ontgen-Research Center for Complex
Material Systems, Universit\"at W\"urzburg, Am Hubland, D-97074 W\"urzburg, Germany}
\author{T. Ala-Nissila}
\affiliation{Department of Mathematical Sciences and Department of Physics, Loughborough University, Loughborough, Leicestershire, LE11 3TU, UK}
\affiliation{COMP Centre of Excellence at the Department of Applied Physics, P.O. Box 11000, FI-00076 Aalto, Finland}
\author{S. Klembt}
\affiliation{Technische Physik, Wilhelm-Conrad-R\"ontgen-Research Center for Complex Material Systems, Universit\"at W\"urzburg, Am Hubland, D-97074 W\"urzburg, Germany}
\author{C. Schneider}
\affiliation{Technische Physik, Wilhelm-Conrad-R\"ontgen-Research Center for Complex
Material Systems, Universit\"at W\"urzburg, Am Hubland, D-97074 W\"urzburg, Germany}
\author{S. H\"ofling}
\affiliation{Technische Physik, Wilhelm-Conrad-R\"ontgen-Research Center for Complex
Material Systems, Universit\"at W\"urzburg, Am Hubland, D-97074 W\"urzburg, Germany}
\affiliation{SUPA, School of Physics and Astronomy, University of St Andrews, St Andrews
KY16 9SS, United Kingdom}

\begin{abstract}
We study the influence of spatial confinement on the second-order temporal coherence of the emission from a semiconductor microcavity in the strong coupling regime. The confinement, provided by etched micropillars, has a favorable impact on the temporal coherence of solid state quasi-condensates that evolve in our device above threshold. By fitting the experimental data with a microscopic quantum theory based on a quantum jump approach, we scrutinize the influence of pump power and confinement and find that phonon-mediated transitions are enhanced in the case of a confined structure, in which the modes split into a discrete set. By increasing the pump power beyond the condensation threshold, temporal coherence significantly improves in devices with increased spatial confinement, as revealed in the transition from thermal to coherent statistics of the emitted light. 
\end{abstract}

\maketitle


\textit{Introduction.---}
The temporal coherence of a source of radiation is a key quantity that distinguishes laser-like devices from thermal emitters. While the first-order coherence function $g^{(1)}(\tau)$, which is the correlator of field amplitudes at different times, reflects the coherence of the emitted photons, its second-order counterpart $g^{(2)}(\tau)$ involves intensity correlation and gives insights into emission statistics. In conventional semiconductor lasers, which rely on stimulated emission and population inversion, the emission statistics transits from thermal (below lasing threshold) to coherent as revealed by $g^{(2)}(\tau=0)=2$ and 1, respectively. 

An example of a light emitting device which may operate as a coherent light source without population inversion is a semiconductor microcavity in the strong light-matter coupling regime \cite{Imamoglu1996}. The strong coupling results in the emergence of hybrid eigenmodes called exciton-polaritons (later, \textit{polaritons})~\cite{Weisbuch1992, KavokinBuch2006}. Being low-mass bosons, they can undergo a dynamic Bose--Einstein condensation (BEC) process at elevated temperatures. Analogous to a conventional laser operating in the weak coupling regime (e.g. the vertical cavity surface emitting laser, VCSEL), the formation of a BEC of polaritons is accompanied by a nonlinear increase of emitted light intensity and a drop in spectral linewidth~ \cite{Imamoglu1996}. The latter is a typical yet not
unambiguous signature of first-order temporal coherence
of the emitted radiation. A more sophisticated approach
relying on Michelson interferometry was discussed in Ref. \cite{Love2008},
where strongly enhanced coherence times in the regime
of polariton lasing were demonstrated by using low-noise
pump sources.

Corresponding to cold atom BECs,
first-order spatial coherence has
been discussed as a key criterion for the claim
of a polariton BEC \cite{Deveaud-Pledran2012, Kasprzak2006, Deng2006, Fischer2014}; however, the second-order
coherence function, representing another important signature
of coherent light, is less well understood. 

From a theoretical point of view, the second order correlation function of the polariton system has been analyzed for the case of weak \cite{Liew.2010, Lemonde.2014, Verger.2006, Flayac.2017} and strong \cite{Demirchyan.2017} coherent resonant pumping.

However, under non-resonant injection, one difficulty arises from the lack of an accurate theoretical quantum description of $g^{(2)}(\tau )$ behavior in real conditions,
such as finite temperatures. Indeed, common methods
are based on stochastic Gross-Pitaevskii \cite{Wouters2009, haug2012temporal, Savenko2013} equations
or expectation value evolution \cite{Savenko2011} that rely on severe
assumptions regarding high-order correlations such
as coherence functions. Additionally, the $g^{(2)}(\tau \neq 0)$ dependence
could not be addressed by semiclassical models.
As a consequence of these complications, reports on
the second-order coherence function in polariton systems
have not been fully conclusive, being rather phenomenological in combined experimental and theoretical works. A second difficulty is based on published experimental results where, in most cases,
a surprisingly slow drop of $g^{(2)}(0)$ above threshold
is commonly observed in two-dimensional (2D) microcavities \cite{Deng2003, Tempel2012, Rahimi-Iman2012} with a persisting value significantly above
unity, thus implying a strong deviation from Poissonian
photon statistics. 

An increase in $g^{(2)}(0)$ with
increasing pump power has even been reported both for GaAs-
and CdTe-based samples \cite{Tempel2012, Kasprzak2008corr, Horikiri2010}. It is reasonable to assume that these peculiarities are the result of a large number of states in the continuous dispersion of polaritons
confined in planar microcavities, which can contribute
to the condensation phenomena and unavoidably
lead to mode competition effects. Such effects further
convolute the coherence phenomena inherent to polariton
condensation. Therefore they are considered as another
main obstacle towards a precise understanding of
the evolution of second-order temporal coherence in solid
state condensates beyond a phenomenological level.

\begin{figure}[tbp]
\centering
\includegraphics[width=\linewidth]{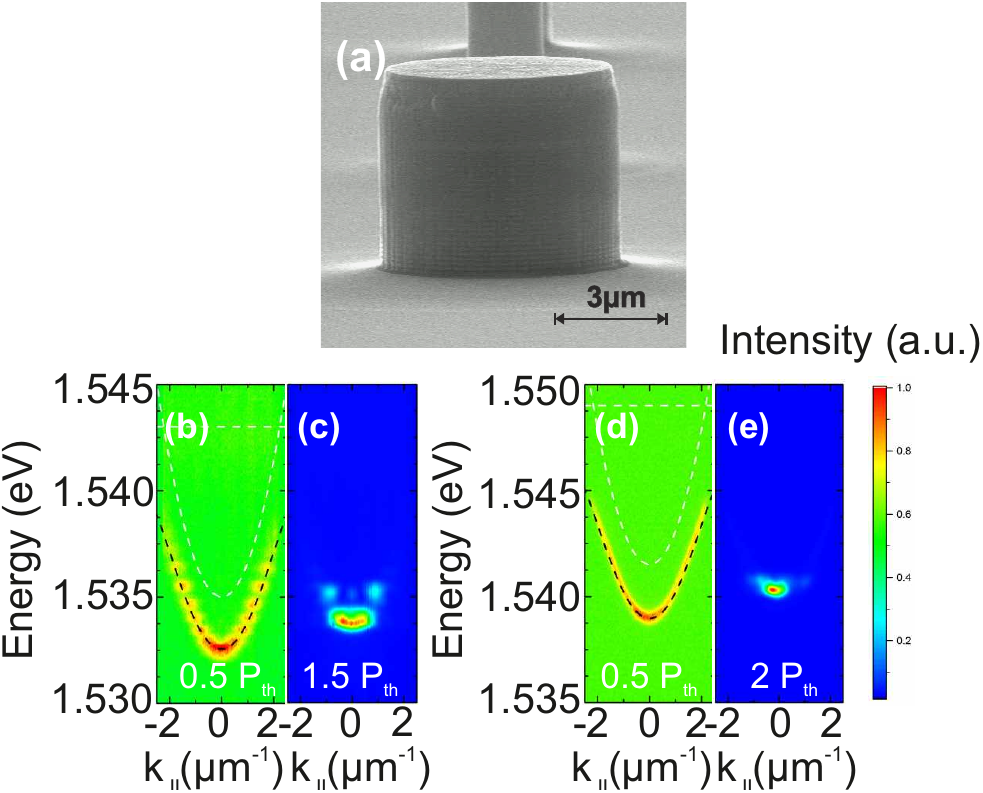}
\caption{(a) Scanning electron microscope image of a processed 6 $\mu$m micropillar under study. (b, c) Energy dispersion measured in photoluminescense (PL) for an 8 $\upmu$m pillar below and above threshold, respectively. (d, e) Emission from the planar sample below and above threshold, respectively. The black dashed lines show the theoretical dispersion from a coupled oscillator model with a detuning of -8/-7.8 meV for the micropillar/planar structure, and the white dashed parabolas and lines show the cavity mode and exciton, respectively.}
\label{Fig1}
\end{figure}

In this Letter, we present a combined experimental and theoretical study of the power dependance of the $g^{(2)}$ function in both confined and planar strongly coupled microcavity structures. Spatial confinement has previously been used to enhance condensation processes in trap structures and micropillars   
 \cite{Dasbach2002, Bajoni2008pil, Klein2015, Schneider2015}; in our case, spatial confinement in 6 to 12 $\mu$m diameter etched micropillars allows us to form polariton
condensates for which $g^{(2)}(0)$ above threshold reaches a plateau with values larger than unity. This $g^{(2)}(0)$ plateau value decreases with tightened optical confinement, reflecting the enhancement of coherence in such structures. When confinement is sufficiently tight to support single-mode condensation, we observe a characterisitc drop of $g^{(2)}(0)$ towards unity with increasing population \cite{Amthor2014, Kim2016}. 
Interestingly, Ref. \cite{Love2008} reports a relatively low $g^{(2)}(0)$ value of 1.1 with a CdTe sample, which can feature significant photonic potential disorder (potential depth of up to 1-2 meV \cite{Krizhanovskii.2009}) and is therefore able to naturally trap the polariton condensate.
Ref. \cite{Kim2016} deals with a sample design based on a hybrid photonic crystal grating to reach tight photonic confinement and is also able to observe a fully coherent state. However, the important role of this energy landscape engineering is neither adressed in a systematic way nor theoretically described in this work.

Our results are supported by a stochastic quantum trajectory approach which provides insight into the interplay between
state occupation, particle 
fluctuations, and photon coherence.
In particular, we show that accounting for dissipative
polariton-phonon interaction and polariton self-interactions
in a fully quantum manner is sufficient to
retrieve the (resolution limited) $g^{(2)}(0)$ behavior in confined structures, where a discrete dispersion assists the
condensation. \\

\textit{Experimental details.---}
The sample under study is a high-Q AlGaAs alloy-based $\lambda/2$ planar microcavity with twelve GaAs quantum wells of 13 nm width, each located in the optical antinodes of the confined electromagnetic field with 23 (27) AlGaAs/AlAs mirror pairs in the top (bottom) distributed Bragg reflector. The Q-factor was experimentally estimated to exceed 12500 for highly photonic structures.

To provide lateral confinement of the polariton modes, we fabricated micropillars with nominal diameters of 6, 8, 10, and 12 $\upmu$m. Electron cyclotron resonance-reactive ion etching was used for a deep etching of the cavity. A scanning electron microscope (SEM) image of a micropillar is shown in Fig.~\ref{Fig1}(a).
We extracted the Rabi-splitting of the device via white light reflection and found a value of 10.1 meV for the heavy hole-based exciton~\cite{Rahimi-Iman2012}.

The sample was pumped with a pulsed Ti:Sa laser tuned to the first Bragg minimum of the stop band located at a wavelength of 749 nm. The pulse width was 2 ps and the nominal resolution of the single-photon detectors was 40 ps, which can be used to extract the second-order correlation function at zero time delay $g^{(2)}(0)$ \cite{Deng2002,Kasprzak2008corr}. The finite length of the emission pulse effectively acts as a time filter in this
configuration \cite{Deng2002}, as only photons emitted from the reservoir within the relaxation time are correlated\cite{Assmann.2010}. 

For excitation beam diameters smaller than the pillar width, we gain high sensitivity of the output to the spatial position of the laser, leading to excitement of the desired optical modes~\cite{kalevich2014controllable}. However, using narrow beams breaks the homogeneity of the pumping scheme, where small pumping spots lead to a repulsive potential that drives the polaritons away from the center (i.e. out of the ground state) thereby causing condensation at high energies and $k$ vectors \cite{Ferrier2011}. To suppress this effect and ensure a homogeneous excitation of the sample, we expanded the beam diameter to $40$  $\mu$m, a value more than three times larger than our largest pillar width. Moreover, to further improve excitation homogeneity, a closed optical aperture and lens were used for beamshaping. \\

\textit{Experimental results and discussion.---}
The optical confinement provided by the micropillar gives rise to a characteristic set of discrete optical modes in the lower polariton dispersion, as seen in Fig.~\ref{Fig1}(b). In contrast, for the case of the planar (unetched) structure, the dispersion has a continuous parabolic shape (Fig.~\ref{Fig1}(d)). Above a distinct threshold occuring at pump powers around 45 W/cm$^2$ ($P_{th}$), we observe the formation of polariton condensate in both the planar and the structured samples (Figs.~\ref{Fig1}(c,e)) (see supplementary informations for details on the input output characteristics).

The temporal second-order correlation function is given as
\begin{equation}
g^{(2)}(\tau)=\frac{\langle I(t+\tau)I(t)\rangle}{\langle I(t) \rangle \langle I(t+\tau \rangle},
\end{equation}
where $I(t)$ is the emission intensity at time t, and $\langle..\rangle$ reads time averaging. For the photon statistics measurements, we use a Hanbury Brown and Twiss configuration with two avalanche photodiodes for single photon counting~\cite{Deng2002,Kasprzak2008corr,Rahimi-Iman2012}. The light emitted from the sample is filtered by a monochromator with a nominal resolution of 0.2 meV. To avoid errors caused by interaction with higher-energy states, special care has been taken to only measure the photons emitted around the energy of the ground state through the use of a spectrometer. 

\begin{figure}[tbp]
\centering
\includegraphics[width=\linewidth]{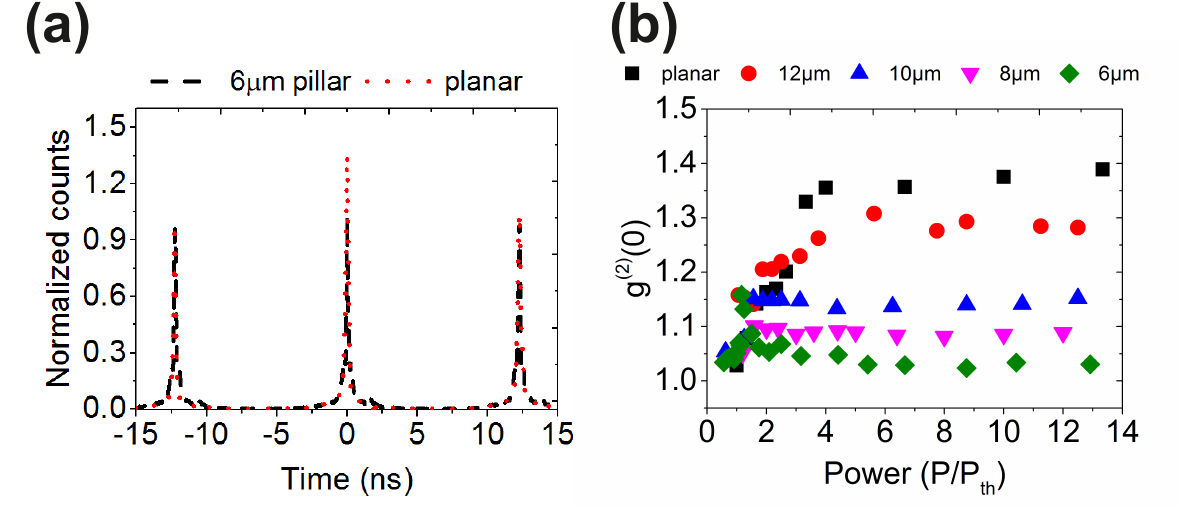}
\caption{(a) Normalized $g^{(2)}(0)$ measurement of the planar sample (red) and the 6$\mu$m pillar sample (black) taken at $7 P_{th}$, displaying the difference in the $g^{(2)}(0)$ value between the confined and 2D system. (b) Comparison between pillars of varying diameter and planar emission. For large pump powers, $g^{(2)}(0)$ reaches a plateau mainly determined by the pillar diameter and thus the confinement of the system.}
\label{Fig2}
\end{figure}

Figure \ref{Fig2}(a) shows the $g^{(2)}(\tau)$ measurements for the planar sample and the smallest investigated pillar (6 $\mu$m). The difference in height of the $\tau=0$ peaks illustrates the difference in the $g^{(2)}(0)$ value at $P=7 P_{th}$. The data has been normalized to the average of the side peak emission.

The increase of the central peak around $\tau = 0$ at and above threshold can be attributed to an enhanced $n>2$-photon emission probability, and indicates that a fully coherent state is not yet formed. To estimate the actual value of $g^{(2)}(0)=N_0/\bar{N}_S$, the integrated intensity of the $\tau=0$ peak, defined as $N_0$, is divided by the average intensity of the $\tau \neq 0$ peaks, $\bar{N}_S=N_s/n$, where $n$ is the number of side peaks and $N_s$ is the integrated side peak intensity.

Figure \ref{Fig2}(b) shows $g^{(2)}(0)$ measurements as a function of pump power for the planar sample and different micropillar sizes. In the case of the 6 $\mu$m micropillar, above threshold the system reaches a nearly coherent state approaching $g^{(2)}(0) = 1$, with a value of $g^{(2)}(0) = 1.03$ at approximately $P=5.5$*$P_{th}$ . For the larger micropillars and the planar structure, $g^{(2)}(0)$ is significantly increased at excitation powers above $P=4$*$P_{th}$, and remains almost constant at a certain value which increases with micropillar size until a maximum in the planar structure. The fact that the polariton system does not reach a fully coherent state of $g^{(2)}(0)=1$ is in qualitative agreement with previous reports~\cite{Deng2002, Tempel2012, Kasprzak2008corr, Rahimi-Iman2012}, where the effect has been assigned to a reservoir depletion triggered by polariton self-scattering from the ground state into higher-energy states. Our simulations indicate that such behavior refers to multimode polariton lasing from the ground state and the existence of a finite set of emitting $k$-values, which lead to an increase in $g^{(2)}(0)$.

Both of these effects should be suppressed in the pillar structures on account of the discrete mode spectrum resulting from optical confinement. In the case of the 6 $\mu$m device, the mode spacing from the ground state to the first excited state amounts to around $0.8$ meV, which drastically reduces the condensate depletion effect at cryogenic temperatures~\cite{Amthor2014}. This allows for the observation of a coherent state in the smallest micropillar device. \\

\begin{figure}[tbp]
\centering
\includegraphics[width=\linewidth]{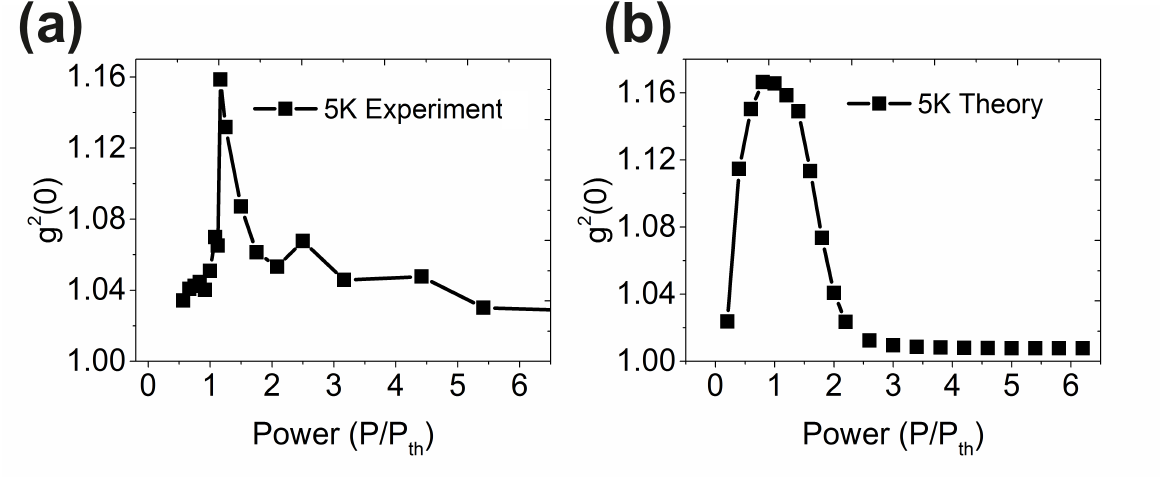}
\caption{(a) Experimental and (b) theoretical $g^{(2)}(0)$ dependence on pump power in the 6 $\mu$m micropillar device, with (b) depicting a fully coherent emission state.}
\label{Fig3}
\end{figure}

\textit{Theoretical description.---}
The coherent interactions are defined by the Hamiltonian
\begin{equation}\label{H}
	\hat{\cal H}= \sum\limits_k {E\left( k_n \right)\hat a_k^\dag {{\hat a}_k} + } \sum\limits_{{k_1},{k_2},p} {{U_{{k_1}{k_2}p}}\hat a_{{k_1}}^\dag \hat a_{{k_2}}^\dag {{\hat a}_{{k_1} + p}}{{\hat a}_{{k_2} - p}}},
\end{equation}
where $\hat a_k$ are bosonic operators for the lower branch polaritons confined by the structure. The condensation occurs in the lower polariton modes, therefore it is necessary to work directly with bosonic polariton operators obtained after a Bogoliubov transformation from the exciton photon fields \cite{Citui.2001}. The mode energies $E\left( k_n \right)$ are set by the dispersion relation
\small
\begin{equation} \label{Ek}
 2E\left( k_n \right) = {{E^{\rm ph}(k_n)} + {E^{\rm ex}(k_n)} \pm \sqrt {{{\left[ {{E^{\rm ph}(k_n)} + {E^{\rm ex}(k_n)}} \right]}^2} + {\Omega_R^2}} }.
\end{equation}
Within the cylindrical geometry imposed by the pillars, the discrete photonic and excitonic energy levels are respectively defined by $E^{\rm ph}(k_n)=\delta+\hbar^2 k_n^2/2m_{\rm ph}$ and $E^{\rm ex}(k_n)=\hbar^2 k_n^2/2m_{\rm ex}$. The cavity photon and exciton effective masses are $m_{\rm ph}=10^{-5} m_0$ and $m_{\rm ex}=0.25 m_0$, respectively, in terms of the free electron mass $m_0$. $k_n$ is the $n^{th}$ zero of the Bessel function $J_0(k,R)$ for a trap of radius $R$, if we assume that the nonresonant pump does not inject angular momentum in the system. $\delta=-8$ meV is the photonic detuning and $\Omega_R=10$ meV is the Rabi splitting. Finally, the interaction term $U_{{k_1}{k_2}p}$ accounts for the polariton-polariton elastic scattering which requires energy-momentum conservation.

The incoherent processes are treated by means of quantum trajectories \cite{Molmer1993}, detailed in Ref. \cite{Flayac2015}, where stochastic quantum jumps are applied to randomly collapse the system wavefunction. This allows to account for the following: (i) Inelastic polariton interactions with a thermal phonon bath of mean population $\bar n_{\rm th}(T)$ following a Bose--Einstein distribution, and in particular the emission or absorbtion of phonons to relax or gain energy with scattering rate $\gamma _{{{\mathbf{k}}_1}{{\mathbf{k}}_2}}^\textrm{ph}$. (ii) The incoherent injection of polaritons with rate $P$ from the exciton reservoir. (iii) Polariton radiative decay at rate $\kappa_k=\hbar/\tau_k$, where $\tau_k$ is the characteristic lifetime (see \cite{Supplement1}). In our simulations, we set an energy cutoff of $E_{\rm max}=4$ meV above the bottom of the polariton dispersion and excite the nearest state with the incoherent pump. By increasing the system diameter, we therefore sample the dispersion with a growing number of states as imposed by Eq. \eqref{Ek}.

The duration of the pulses absorbed by the cavity is significantly enhanced due to the interaction with the long-living exciton reservoir, which extends over several polariton lifetimes. Therefore, to reduce the computational cost required by a two-time analysis of the second-order correlations required by a pulsed excitation scheme, we assume the the system has time to reach its steady-state statistics along the pulse. To mimic the Hanbury Brown and Twiss setup, we then record the quantum jump events associated with radiative decays and reconstruct the delayed correlations \cite{Flayac2014}
\begin{equation}\label{g2k}
  g_k^{\left( 2 \right)}\left( \tau  \right) = \frac{{\langle {\hat a_k^\dag \left( 0 \right)\hat a_k^\dag \left( \tau  \right){{\hat a}_k}\left( \tau  \right){{\hat a}_k}\left( 0 \right)} \rangle }}{{\langle {\hat a_k^\dag \left( \tau  \right){{\hat a}_k}\left( \tau  \right)} \rangle \langle {\hat a_k^\dag \left( 0 \right){{\hat a}_k}\left( 0 \right)} \rangle }}.
\end{equation}
Then, the equal time correlations accounting for the finite temporal resolution of the detector $T_{\rm res}$ are obtained as
\begin{equation}\label{g2res}
  g_{k,\rm  res}^{\left( 2 \right)}\left( 0 \right) = \frac{{\int\limits_0^{{T_{\rm res}}} {{G_k^{(2)}}\left( \tau  \right)d\tau } }}{{\int\limits_0^{{T_{\rm res}}} {n_k\left( \tau  \right)n_k\left( 0 \right)d\tau } }},
\end{equation}
where ${{G_k^{(2)}}\left( \tau  \right)}$ is the numerator of Eq. \ref{g2k} and $n_k(\tau)=\langle\hat a_k^\dag\hat a_k\rangle$.

Figure \ref{Fig3}(a) displays the experimental results of $g^{(2)}(0)$ evolution in the 6 $\mu$m device. In Fig. \ref{Fig3}(b), we show the theoretical results obtained for the ground state $g_{0,\rm  res}^{\left( 2 \right)}\left( 0 \right)$ function in the case of the 6 $\mu$m pillar at 5 K; in principal, we theoretically reproduce the rise and fall of the correlation functions versus the pump power. In particular, the initial rise in the coherence function has not been previously modeled \cite{Schwendimann2008}. Such a feature emerges within our approach from the photon counting process. Indeed, below threshold, emission events are rare, making the system evolution quasi-unitary, and therefore the system essentially displays Poissonian statistics. This behavior cannot be reproduced by a master equation treatment that predicts a thermal $g_2$=2 value. The statistics then reveal a competition between coherence and thermalization at the peak until the condensation threshold is reached, and coherence is established.

\begin{figure}[tbp]
\centering
\includegraphics[width=0.5\columnwidth]{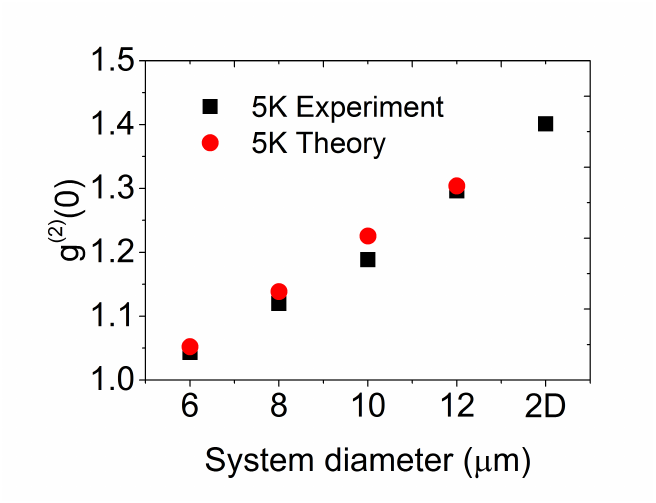}
\caption{Experimental and theoretical values of $g^{(2)}(0)$ exhibiting the continuous increase of temporal coherence with increased system size for a comparable emitted groundstate intensity at pump powers far above threshold for all systems.}
\label{Fig4}
\end{figure}

Figure \ref{Fig4} depicts the combined experimental and theoretical results of the $g^{(2)}(0)$ value far above condensation threshold for a comparable output intensity of the ground state for the series of micropillar devices and the 2D planar reference structure. We reveal a the dependence of temporal coherence on confinement by observing a significant drop in $g^{(2)}(0)$ with system size. Further, we were able to theoretically recreate the increase in the second-order correlation function versus the system size above threshold. Such features mainly result from the nontrivial polariton-phonon scattering rate dependence with respect to state separation (see \cite{Supplement1}). Coherence is favored for smaller pillars, where relaxation towards the ground state is consequently more efficient. Note that we have considered the multimode emission here by computing the second-order correlation $g_{\rm out}^{\left( 2 \right)}\left( 0 \right)$ over the total output field $\hat a_{\rm out}=\sum\nolimits_k {\sqrt {{\kappa _k}} {{\hat a}_k}}+\hat a_{\rm in}$ following the input-output theory \cite{Collett1984}. These results underline the power of the quantum jump approach for a discrete polariton mode scheme, and support the conclusion that the presence of optical confinement strongly improves the coherence of our polariton laser. \\

\textit{Conclusion.---}

We have evaluated the effect of optical confinement on the temporal coherence function behavior of a polariton condensate. Experimentally, we observed a significantly decreased $g^{(2)}(\tau=0)$ value above the polariton lasing threshold in tightly confined systems, as compared to more planar polariton microcavity reference stuctures. By combining our experimental data with a microscopic model based on the quantum jump approach, we were able to successfully describe the coherence evolution in our polariton devices, which is of paramount importance for the design of next-generation coherent polariton light sources.  This beneficial effect of mode localization can further serve in other open dissipative quantum systems like photonic condensates \cite{Klaers.2010}, strongly localized excitons dressed to cavity fields \cite{Dietrich.2016} as well as the entire community working on semiconductor lasers.  \\

\section*{Acknowledgments}
The authors would like to thank the State of Bavaria for financial support. We thank Monika Emmerling and Adriana Wolf for expert sample processing and Joel Rasmussen (RECON) for a critical reading of our manuscript.. T.A-N. has been supported in part by the Academy of Finland through its CoE grants 251748 and 284621.
I.G.S. acknowledges the support of the Institute for Basic Science in Korea (project IBS-R024-D1), the Australian Research Council’s Discovery Projects funding scheme (project DE160100167), Russian Science Foundation (project 17-12-01039) and President of Russian Federation (project МК-5903.2016.2). Funding by the EPSRC within the Hybrid Polaritonics grant (EP/M025330/1) is gratefully acknowledged. C.S. and M.K. thank the DFG within the project Schn1376-3.1. Funding by the EPSRC within the Hybrid Polaritonics grant (EP/M025330/1) is gratefully acknowledged. S.K. acknowledges the European Commission for the H2020 Marie Skłodowska-Curie Actions (MSCA) fellowship (Topopolis).


\bibliographystyle{osajnl}

\end{document}